\documentstyle[prl,eqsecnum,aps]{revtex}

\begin{document}
\author{Jian-Qi Shen  \footnote{E-mail address: jqshen@coer.zju.edu.cn}}
\address{Zhejiang Institute of Modern Physics and Department of
Physics, Zhejiang University, Hangzhou 310027, P.R. China}
\date{\today }
\title{The Connection Between Density Matrix Method, Supersymmetric Quantum Mechanics and Lewis-Riesenfeld Invariant Theory\footnote{This paper was initiated three years ago (in 2000).}}
\maketitle
\begin{abstract}
This paper is concerned with the connection between density matrix method, supersymmetric 
quantum mechanics and Lewis-Riesenfeld invariant theory. It is shown that these three formulations share 
the common mathematical structure: specifically, all of them have the invariant operators which satisfies the Liouville-Von Neumann
equation and the solutions to the time-dependent Schr\"{o}dinger equation and/or Schr\"{o}dinger eigenvalue equation can be constructed in terms of the eigenstates of the invariants.

\end{abstract}
\section{Lewis-Riesenfeld invariant theory}
Lewis-Riesenfeld invariant theory can be applied to the solutions of time-dependent Schr\"{o}dinger equation.  

In order to illustrate the Lewis-Riesenfeld invariant theory\cite{Lewis} easily, we
consider a one-dimensional system whose Hamiltonian $H(t)$ is
time-dependent. According to the Lewis-Riesenfeld invariant theory\cite{Lewis}, a Hermitian operator $I(t)$ 
is called invariant if it satisfies the following invariant equation, {\it i.e.}, the Liouville-Von Neumann
equation (in the unit $\hbar =1$)

\begin{equation}
\frac{\partial I(t)}{\partial t}+\frac{1}{i}[I(t),H(t)]=0.  \label{eq21}
\end{equation}
The eigenvalue equation of the time-dependent invariant $\left| \lambda
_{n},t\right\rangle $ is given
\begin{equation}
I(t)\left| \lambda _{n},t\right\rangle =\lambda _{n}\left| \lambda
_{n},t\right\rangle,  \label{eq22}
\end{equation}
where
\begin{equation}
\frac{\partial \lambda _{n}}{\partial t}=0.  \label{eq23}
\end{equation}
The time-dependent Schr\"{o}dinger equation (in the unit $\hbar =1$) for the
system is

\begin{equation}
i\frac{\partial \left| \Psi (t)\right\rangle _{s}}{\partial t}=H(t)\left|
\Psi (t)\right\rangle _{s}.  \label{eq24}
\end{equation}
In terms of the Lewis-Riesenfeld invariant theory, the particular solution $\left|
\lambda _{n},t\right\rangle _{s}$ of Eq.(\ref{eq24}) differs from the
eigenfunction $\left| \lambda _{n},t\right\rangle $ of the invariant $I(t)$
only by a phase factor $\exp [\frac{1}{i}\phi _{n}(t)]$, then the general
solution of the Schr\"{o}dinger equation (\ref{eq24}) can be written as

\begin{equation}
\left| \Psi (t)\right\rangle _{s}=%
\mathop{\textstyle\sum}%
_{n}C_{n}\exp [\frac{1}{i}\phi _{n}(t)]\left| \lambda _{n},t\right\rangle ,
\label{eq25}
\end{equation}
where

\[
\phi _{n}(t)=\int_{0}^{t}\left\langle \lambda _{n},t^{^{\prime
}}\right| H(t^{^{\prime }})-i\frac{\partial }{\partial t^{^{\prime
}}}\left| \lambda _{n},t^{^{\prime }}\right\rangle {\rm
d}t^{^{\prime }},
\]

\begin{equation}
C_{n}=\langle \lambda _{n},t=0\left| \Psi (t=0)\right\rangle _{s}.
\label{eq26}
\end{equation}
$\left| \lambda _{n},t\right\rangle _{s}=\exp [\frac{1}{i}\phi
_{n}(t)]\left| \lambda _{n},t\right\rangle $ $(n=1,2,\cdots)$ are said to
form a complete set of the solutions of Eq.(\ref{eq24}). The statement
outlined above is the basic content of the Lewis-Riesenfeld invariant theory.

A time-dependent unitary transformation operator can be constructed to
transform $I(t)$ into a time-independent invariant $I_{V}\equiv V^{\dagger
}(t)I(t)V(t)$ \cite{Gao1,Gao3} with
\begin{eqnarray}
I_{V}\left| \lambda _{n}\right\rangle &=&\lambda _{n}\left| \lambda
_{n}\right\rangle ,  \label{eq27} \\
\left| \lambda _{n}\right\rangle &=&V^{\dagger }(t)\left| \lambda
_{n},t\right\rangle .  \label{eq28}
\end{eqnarray}
Under the unitary transformation $V(t),$ the Hamiltonian $H(t)$ is
correspondingly changed into $H_{V}(t)$

\begin{equation}
H_{V}(t)=V^{\dagger }(t)H(t)V(t)-V^{\dagger }(t)i\frac{\partial V(t)}{%
\partial t}.  \label{eq29}
\end{equation}
In accordance with this unitary transformation method\cite{Gao1}, it is very
easy to verify that the particular solution $\left| \lambda
_{n},t\right\rangle _{s0}$ of the time-dependent Schr\"{o}dinger equation
associated with $H_{V}(t)$

\begin{equation}
i\frac{\partial \left| \lambda _{n},t\right\rangle _{s0}}{\partial t}%
=H_{V}(t)\left| \lambda _{n},t\right\rangle _{s0}  \label{eq210}
\end{equation}
is different from the eigenfunction $\left| \lambda _{n}\right\rangle $ of $%
I_{V}$ only by the same phase factor $\exp [\frac{1}{i}\phi _{n}(t)]$ as
that in Eq.(\ref{eq25}), {\it i.e.},

\begin{equation}
\left| \lambda _{n},t\right\rangle _{s0}=\exp [\frac{1}{i}\phi
_{n}(t)]\left| \lambda _{n}\right\rangle .  \label{eq211}
\end{equation}
Substitution of $\left| \lambda _{n},t\right\rangle _{s0}$ of Eq.(\ref
{eq210}) into Eq.(\ref{eq211}) yields

\begin{equation}
\dot{\phi}(t)\left| \lambda _{n}\right\rangle =H_{V}(t)\left| \lambda
_{n}\right\rangle ,  \label{eq212}
\end{equation}
which means that $H_{V}(t)$ differs from $I_{V}(t)$ only by a time-dependent
multiplying c-number factor. It can be seen from Eq.(\ref{eq212}) that the
particular solution of Eq.(\ref{eq210}) can be easily obtained by
calculating the phase from Eq.(\ref{eq212}). Thus, one is led to the
conclusion that if the $V(t),$ $I_{V},$ $H_{V}(t)$ and the eigenfunction $%
\left| \lambda _{n}\right\rangle $ of $I_{V}$ have been found, the problem
of solving the complicated time-dependent Schr\"{o}dinger equation (\ref
{eq24}) reduces to that of solving the much simplified equation (\ref{eq210}%
). 
\section{Density matrix method}                             
Density matrix method has many helpful applications to the semiclassical theory of laser\cite{Li}. 
     
The density matrix operator is defined to be $\rho(t)=|\Psi(t)\rangle\langle\Psi(t)|$, 
the wavefunction $|\Psi(t)\rangle$ of which agrees with the time-dependent Schr\"{o}dinger equation.
It follows that the density operator satisfies the following time-evolution equation
\begin{equation} 
\frac{\partial \rho(t)}{\partial t}+\frac{1}{i}[\rho(t),H(t)]=0,  \label{eq31}   
\end{equation} 
which is just the Liouville-Von Neumann
equation (\ref{eq21}). This, therefore,means that the density operator $\rho(t)$ is the Lewis-Riesenfeld invariant.

In what follows we will discuss the density matrix method that solves the time-dependent
Schr\"{o}dinger equation governing the semiclassical laser process. As an illustrative example, 
we consider a two-level atomic system ($|a\rangle$ and $|b\rangle$) interacting with the classical Maxwellian electromagnetic fields. 
The time-dependent wavefunction $|\Psi(t)\rangle$ can be expressed in terms of $|a\rangle$ and $|b\rangle$, namely,
\begin{equation}
|\Psi(t)\rangle=c_{a}(t)|a\rangle+c_{b}(t)|b\rangle,         \label{eq}
\end{equation}   
and the density matrix can be rewritten as
\begin{equation}     
\rho(t)=\left(\begin{array}{cc} c_{a} \\
c_{b} \end {array}\right)(c_{a}^{\ast} \quad c_{b}^{\ast})
=\left(\begin{array}{cc}
c_{a}c_{a}^{\ast} & c_{a}c_{b}^{\ast} \\
c_{b}c_{a}^{\ast} & c_{b}c_{b}^{\ast}               \label{eq32}
\end{array}
\right)
\end{equation}     
in the representation of $|a\rangle$ and $|b\rangle$. 

The Hamiltonian of two-level atomic system takes the form
 \begin{equation} 
 H=H_{0}+V,
 \end{equation}    
where the atomic free Hamiltonian $H_{0}$ and the interaction Hamiltonian $V$ are respectively of the form
\begin{equation} 
H_{0}=\left(\begin{array}{cc}
\omega_{a} & 0 \\
0 &  \omega_{b}
\end{array}
\right),      \quad   V=\left(\begin{array}{cc}
0 & V_{ab} \\
V_{ba} & 0 
\end{array}
\right).                \label{eq33}                        
\end{equation} 
Substitution of (\ref{eq32})-(\ref{eq33}) into (\ref{eq31}) yields
\begin{equation}
\dot{\rho_{aa}}=-[iV_{ab}\rho_{ba}+{\rm c.c}],  \quad       \dot{\rho_{bb}}=iV_{ab}\rho_{ba}+{\rm c.c}, \quad      \dot{\rho_{ab}}=-i(\omega_{a}-\omega_{b})\rho_{ab}+iV_{ab}(\rho_{aa}-\rho_{bb})   \label{eq37}
\end{equation}
with  $\rho_{aa}=c_{a}c_{a}^{\ast}$, $\rho_{ab}=c_{a}c_{b}^{\ast}$, $\rho_{ba}=c_{b}c_{a}^{\ast}$ and $\rho_{bb}=c_{b}c_{b}^{\ast}$.   
So the resolution of the Schr\"{o}dinger equation can be ascribed to the solution of the Eq.(\ref{eq37}) ({\it i.e.}, the Liouville-Von Neumann
equation (\ref{eq21})), which is similar to the case in Lewis-Riesenfeld invariant theory, where we should obtain the expression for the invariant $I(t)$ via the Liouville-Von Neumann
equation and solve the eigenvalue equation (\ref{eq22}) of $I(t)$.  
\section{Supersymmetric quantum mechanics}
Supersymmetric quantum mechanics is applicable to the energy eigenvalue problems of one-dimensional potential walls. 

The sequence of ideas of supersymmetric quantum mechanics is to obtain the eigenstates of the following two Hamiltonians
\begin{equation}
H_{\pm}=-\frac{\hbar^{2}}{2m}\frac{{\rm d}^{2}}{{\rm d}x^{2}}+V_{\pm}(x)
\end{equation}
via the eigenstates of $H=-\frac{\hbar^{2}}{2m}\frac{{\rm d}^{2}}{{\rm d}x^{2}}+V(x)$, where $V_{\pm}(x)$ agrees with the following two equations:
\begin{equation}
\frac{1}{2}[V_{+}(x)+V_{-}(x)]=W(x)^{2},    \quad     \frac{1}{2}[V_{+}(x)-V_{-}(x)]=\frac{\hbar}{\sqrt{2m}}W'(x)     \label{41}
\end{equation}       
with $W'(x)$ denoting the derivative of $W(x)$ with respect to $x$, and $W(x)$ is so defined that the ground state $\psi_{0}(x)$ of $H$ is 
\begin{equation}
\psi_{0}(x)=\exp\left[-\frac{\sqrt{2m}}{\hbar}\int^{x}{\rm d}xW(x)\right].
\end{equation}   
Define the lowering and raising operators $A$ and $A^{\dagger}$ as follows
\begin{equation}
A=\frac{\hbar}{\sqrt{2m}}\frac{{\rm d}}{{\rm d}x}+W(x),    \quad        A^{\dagger}=-\frac{\hbar}{\sqrt{2m}}\frac{{\rm d}}{{\rm d}x}+W(x),
\end{equation}       
and therefore we have $H_{-}=A^{\dagger}A$ and $H_{+}=AA^{\dagger}$ and the commuting relation $[A,A^{\dagger}]=\frac{2\hbar}{\sqrt{2m}}W'(x)$.

By using the properties of lowering and raising operators $A$ and $A^{\dagger}$, we can obtain the eigenstates of $H_{\pm}$, which are constructed in terms of the eigenstates of $H$. 

It is readily verified that here the Hamiltonian $H$ serves as the invariant that satisfies the Liouville-Von Neumann
equation. Since here $H$ is {\it time-independent}, we should calculate the commuting relation $[H_{+},H]$ only.
According to Eqs.(\ref{41}), we obtain $V_{+}=W(x)^{2}+\frac{\hbar}{\sqrt{2m}}W'(x)$ and consequently 
\begin{equation}
V_{+}=\frac{\hbar^{2}}{2m}\frac{\frac{{\rm d}^{2}\psi_{0}(x)}{{\rm d}x^{2}}}{\psi_{0}(x)}. 
\end{equation}        
In accordance with the eigenvalue equation $[-\frac{\hbar^{2}}{2m}\frac{{\rm d}^{2}}{{\rm d}x^{2}}+V(x)]\psi_{0}(x)=\varepsilon_{0}\psi_{0}(x)$, we get $V_{+}(x)=V(x)-\varepsilon_{0}$. So 
$[H_{+},H]=[V_{+}(x)-V(x), -\frac{\hbar^{2}}{2m}\frac{{\rm d}^{2}}{{\rm d}x^{2}}]=0$, because of $V_{+}(x)-V(x)=-\varepsilon_{0}$. This, therefore, means that the Hamiltonian $H$ is an invariant.   

For the detailed applications of supersymmetric quantum mechanics to the energy eigenvalue problems of one-dimensional potential walls and related problems, Readers may be referred to the references \cite{Cooper,Gendenshtein,Sukumar,Dutt}. 
\section{Concluding Remarks}
In general, the Lewis-Riesenfeld invariant theory is applicable to the multi-level atomic quantum system interacting with the second-quantized electromagnetic fields\cite{Shen1,Shen2}, 
while the density matrix method is appropriate to treat the multi-level atomic quantum system interacting with the classical Maxwellian electromagnetic fields (waves)\cite{Li}. The former formulation can also be applied to 
the motion of a charged spinning particle in a classical magnetic fieldcite\cite{Shen3} or of a photon moving in a sufficiently perfect curved fibercite\cite{Shen1}. 

In the present paper we show that the above three formulations share 
the common mathematical structure: specifically, all of them have the invariant operators which satisfies the Liouville-Von Neumann
equation and the solutions to the time-dependent Schr\"{o}dinger equation and/or Schr\"{o}dinger eigenvalue equation can be constructed in terms of the eigenstates of the invariants.   
\\ \\
\textbf{Acknowledgements}  The author was grateful to X.C. Gao for useful
suggestions.

\end{document}